%% file: ProceedingsLHCP2017_StrongSUSY_clseitz_vOct13_arxiv.tex
\documentclass[10pt]{article}
\usepackage{graphicx}

\usepackage{hyperref}
\def\Title#1{\begin{center} {\Large #1 } \end{center}}
\def\Author#1{\begin{center}{ \sc #1} \end{center}}
\def\Address#1{\begin{center}{ \it #1} \end{center}}

\newcommand\pubblock{\rightline{\begin{tabular}{l} Proceedings of the Fifth Annual LHCP\\ \pubnumber\\
         \pubdate  \end{tabular}}}

\newenvironment{Abstract}{\begin{quotation} \begin{center} 
             \large ABSTRACT \end{center}\bigskip 
      \begin{center}\begin{large}}{\end{large}\end{center} \end{quotation}}

\newenvironment{Presented}{\begin{quotation} \begin{center} 
             PRESENTED AT\end{center}\bigskip 
      \begin{center}\begin{large}}{\end{large}\end{center} \end{quotation}}

\input econfmacros.tex

\usepackage{color}

\newcommand\pubnumber{  }

\newcommand\pubdate{\today}

\def\affiliation{
On behalf of the ATLAS and CMS Collaborations, \\
Physik-Institut\\
 University of Z\"urich, Winterthurerstrasse  190, 8057 Z\"urich,Switzerland}

\newcommand{\fbinv}{\ensuremath{\mathrm{fb}^{-1}}}
\newcommand{\LT}{\ensuremath{L_\mathrm{T}}}
\newcommand{\pT}{\ensuremath{p_\mathrm{T}}}
\newcommand{\HT}{\ensuremath{H_\mathrm{T}}}
\newcommand{\MHT}{\ensuremath{H_\mathrm{T}^{miss}}}
\newcommand{\MpT}{\ensuremath{p_\mathrm{T}^{miss}}}
\newcommand{\MpTHT}{\ensuremath{p_\mathrm{T}^{miss}/\sqrt{H_T}}}
\newcommand{\meff}{\ensuremath{m_\mathrm{eff}}}
\newcommand{\njet}{\ensuremath{N_\mathrm{Jet}}}
\newcommand{\ttbar}{\ensuremath{t\bar{t}}}

\usepackage{cite}

\begin{document}

\large
\begin{titlepage}
\pubblock

\vfill
\Title{ Searches for strong production of supersymmetry at ATLAS and CMS }
\vfill

\Author{ Claudia Seitz  }
\Address{\affiliation}
\vfill
\begin{Abstract}
Searches for physics beyond the standard model are considered one of the corner stones of the physics program at the two multipurpose experiments, ATLAS and CMS, at the LHC.
Both experiments have conducted numerous searches for new supersymmetric particles in a wide variety of final states at a center-of-mass energy of $\sqrt{s} = 13$~TeV. The datasets discussed in these
proceedings consist of 36~fb$^{-1}$ of proton-proton collisions collected during the 2016 run of the LHC. No signs of new physics are found, however, limits on different supersymmetric scenarios are placed and
extend previous exclusion regions from 7 and 8 TeV by several hundred GeV.
\end{Abstract}
\vfill

\begin{Presented}
The Fifth Annual Conference\\
 on Large Hadron Collider Physics \\
Shanghai Jiao Tong University, Shanghai, China\\ 
May 15-20, 2017
\end{Presented}
\vfill
\end{titlepage}
\def\thefootnote{\fnsymbol{footnote}}
\setcounter{footnote}{0}
%

\normalsize 


\section{Introduction}

The standard model (SM) of particle physics is one of the most successful theories developed during the last century.
Despite its extreme success there are still many unanswered questions, most of which require an extension to the existing theory.
 One of the favored extensions of the SM is supersymmetry (SUSY)~\cite{Ramond:1971gb,Golfand:1971iw,Neveu:1971rx,                                                                                                               
Volkov:1972jx,Wess:1973kz,Wess:1974tw,Fayet:1974pd,Nilles:1983ge}, predicting a variety of new particles at the TeV scale, that differ in spin from their SM counterparts.
 SUSY provides an elegant solution to some of the shortcomings of the SM. For instance, due to the presence of new supersymmetric particles with different spin (i.e. fermionic partners to SM bosons and vice versa) cancelations are introduced in the corrections to the Higgs mass, which can stabilize it at the measured value of about 125~GeV~\cite{Aad:2015zhl}.  Introducing a new quantum number called $R$-parity~\cite{FARRAR1978575} and assuming it is a conserved quantity postulates that SUSY particles are produced in pairs and their decays contain in the end the lightest SUSY particle (LSP). This particle would escape detection and presents an excellent
 dark matter candidate. As an additional part of the new theory, it is possible to unify the gauge couplings of the electromagnetic, weak, and strong force at an energy close to the Planck scale.
 
Searches for physics beyond the SM are considered one of the corner stones of the physics program at the two multipurpose experiments, ATLAS~\cite{Aad:2008zzm} and CMS~\cite{Chatrchyan:2008aa}, at the LHC.
Both experiments have conducted numerous searches for new supersymmetric particles in a wide variety of final states at a center-of-mass energy of $\sqrt{s} = 13$~TeV. The datasets discussed in these
proceedings consist of 36~\fbinv~of proton-proton collisions collected during the 2016 run of the LHC. 
The results are interpreted in terms of simplified models, where only a small number of particles are considered accessible at the current center-of-mass energy. Specifically, in these proceedings scenarios of strong production of gluinos and light-flavor squarks are considered leading to multiple different final states.
For all discussed scenarios we assume the conservation of $R$-parity, therefore leading to final states where at least two LSPs escape detection. The expected signal production cross sections are computed at NLO with next-to-leading-logarithm (NLL)
accuracy~\cite{bib-nlo-nll-01,bib-nlo-nll-02,bib-nlo-nll-03,bib-nlo-nll-04,bib-nlo-nll-05}.
No sign of new physics is found and the results are typically interpreted as upper limits at the 95\% confidence level on the production cross section within a two dimensional mass plane of the strongly coupled SUSY particle and the LSP. 
\begin{figure}[htb]
\centering
\includegraphics[width=0.27\textwidth]{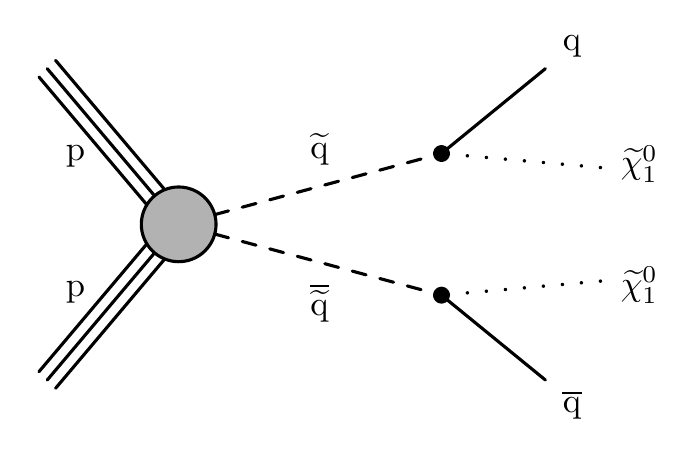}
\includegraphics[width=0.27\textwidth]{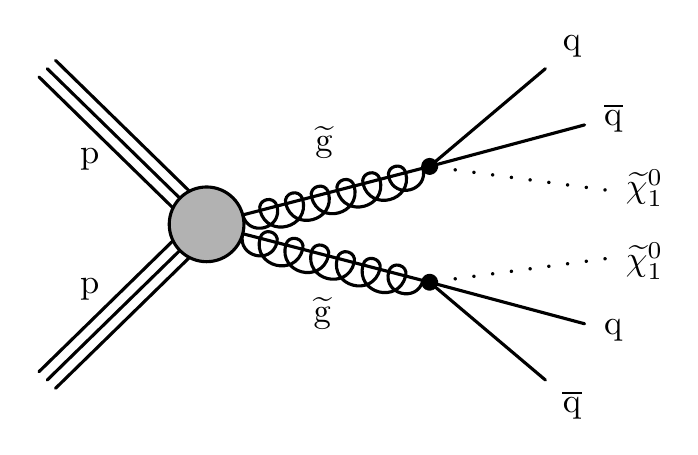}
\includegraphics[width=0.27\textwidth]{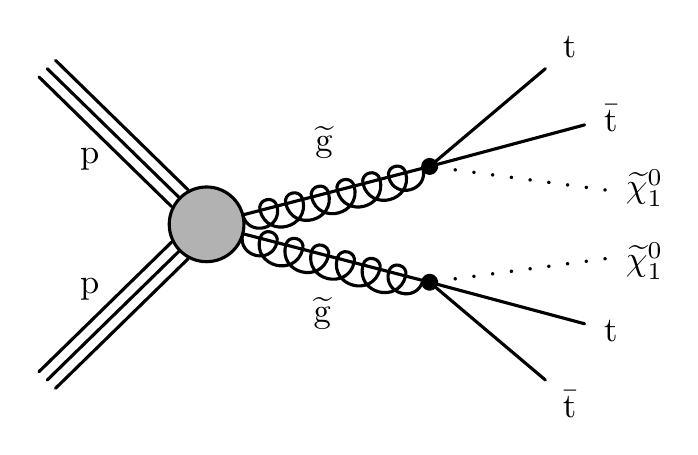}\\
\includegraphics[width=0.21\textwidth]{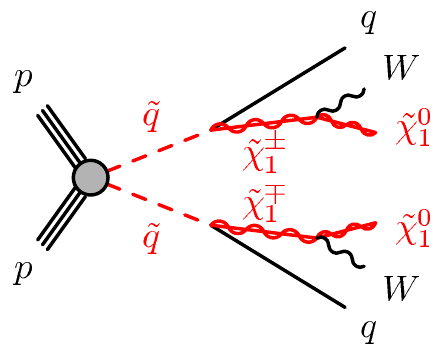}
\includegraphics[width=0.21\textwidth]{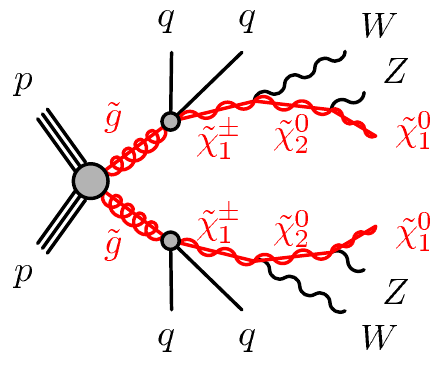}
\caption{Simplified model diagrams. Shown are the pair production and direct decay of light flavor squarks (top, left) and the pair production
of gluinos with direct decays into light flavor quarks (top, middle) and top quarks (top, right) and the LSP.
Additionally, one-step cascade decays of light flavor squarks that include W or Z bosons in the the decay (bottom, left) and two-step
cascade decays of gluinos that include W and Z bosons (bottom, right) are depicted as well.}
\label{fig:feynman}
\end{figure}

The report itself discusses these searches by separating them into different number of leptons in the final states. 
The following Sections 2 and 3 describe searches with zero leptons and up to one lepton, respectively. Section 4 focuses on multilepton final states.
We conclude the report with a summary given in Section 5.

\section{Fully hadronic searches}
\subsection{Analyses strategy}
Several different searches are performed by CMS~\cite{CMS-SUS-16-033, CMS-SUS-16-036} and ATLAS~\cite{ATLAS-2017-022, ATLAS-2017-33} in the fully hadronic final state, i.e. where events containing leptons are vetoed.
Two searches for strong production of gluinos and squarks conducted by CMS are based on multiple exclusive search regions, that make use of different kinematic variables. The analysis presented in Ref.~\cite{CMS-SUS-16-033} defines these exclusive search regions in terms of the number of jets \njet, the number of b jets,
the scalar sum \HT~of the transverse momenta \pT~of the jets, and the magnitude \MHT~of the vector \pT~sum of the jets.
Complementary to this, the analysis described in Ref.~\cite{CMS-SUS-16-036} makes use of the MT2 variable~\cite{LESTER199999} in addition to exclusive regions in the number of jets, the number of b jets, and \HT.
In particular the MT2 variable is generalized for events with at least two jets by clustering these reconstructed jets into two pseudo-jets and then calculate MT2 as described in ~\cite{CMS-SUS-16-036} using the jets and the missing transverse momentum of the event \MpT.
Both searches follow an inclusive strategy, meaning they target many of the aforementioned simplified models.
Lower jet and b jet multiplicity final states, targeting light squark production (Fig.~\ref{fig:feynman} top, left), are included as well as high jet and b jet multiplicity ones, targeting the striking signature of gluino pair production with up to four top quarks in the final state (Fig.~\ref{fig:feynman} top, right).

\begin{figure}[htb]
\centering
\includegraphics[width=0.4\textwidth]{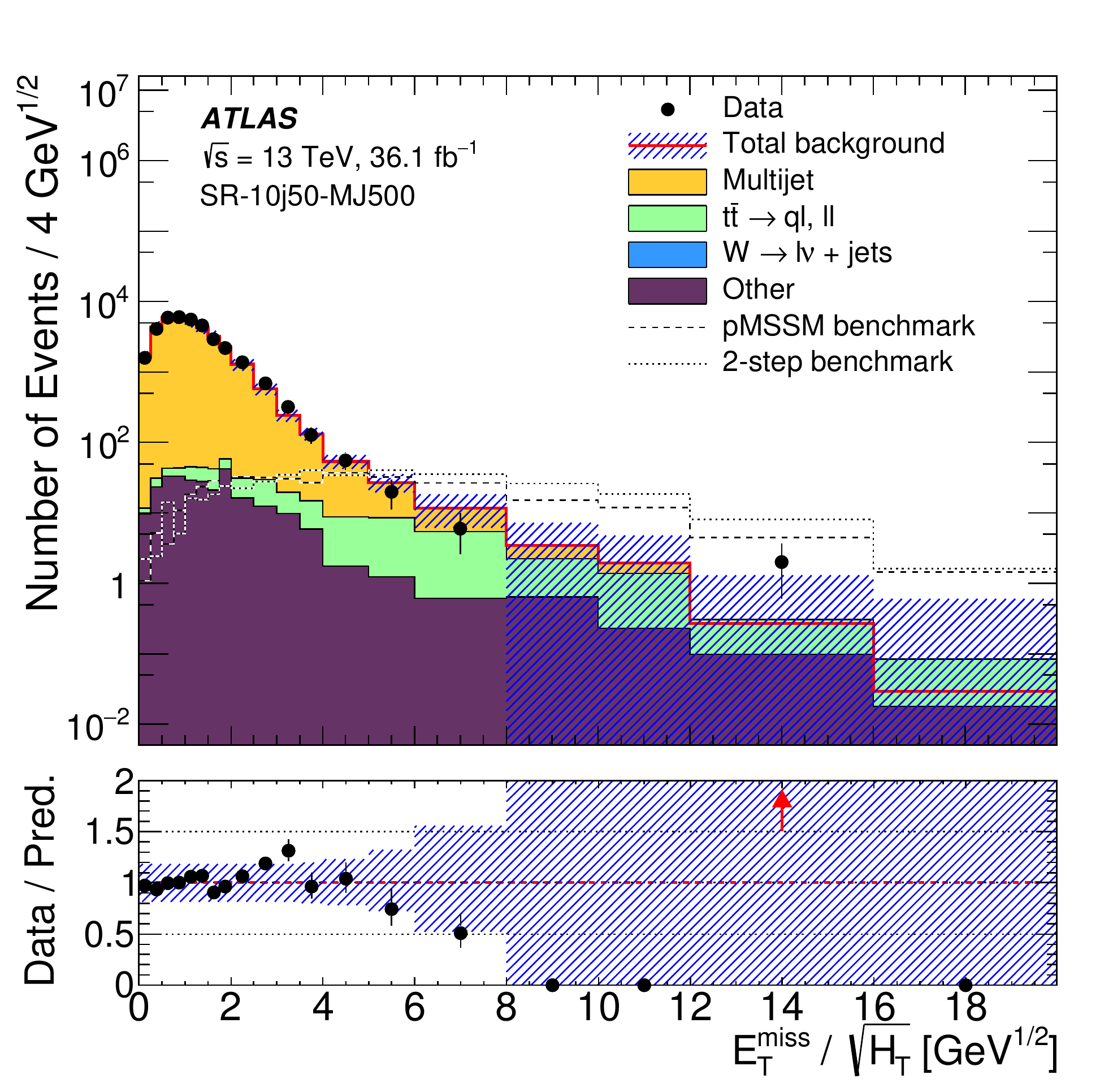}
\caption{Search region from Ref.~\cite{ATLAS-2017-33}. The plot shows one of the \MpTHT~distribution in a search regions of the analysis along with the background prediction.}
\label{fig:regionresult17_033}
\end{figure}

Equivalent simplified models are targeted by the ATLAS searches. While the analysis discussed in Ref.~\cite{ATLAS-2017-022} focuses on low jet multiplicity topologies without heavy flavor in the decay, Ref.~\cite{ATLAS-2017-33} targets higher jet multiplicity final states that include cascade decays into W and Z bosons, as shown in Fig.~\ref{fig:feynman} bottom, right.
Both complementary searches rely on the definition of inclusive search regions which are optimized for different simplified models and mass scenarios to cover a large possible phase space. 
Two separate approaches are used in the low \njet~selection. The first one is the Recursive Jigsaw
Reconstruction (RJR) technique~\cite{PhysRevD.95.035031}, where specific assumptions on the underlying decay topology are made and used to calculate additional higher level discriminating variables. The second approach is based on the effective mass variable \meff, which represents the sum \pT~of all reconstructed objects plus \MpT. Additional requirements on the number of small radius R$=$0.4 jets as well as large radius R$=$1.0 jets are used in defining the search regions.
The search covering higher jet multiplicity final states relies on the discriminating variables of the mass of large radius jets $MJ$ as well as the distinct shape of the \MpTHT~distribution for QCD multijet events  when compared with signal events, as shown in Fig.~\ref{fig:regionresult17_033}. Additional requirements on the minimum \pT~of the selected jets as well as the number of b jets are used to define the search regions.

\subsection{Background modeling}
The main backgrounds for these fully hadronic final states depend largely on the search regions, for example the number of jets or b jets, however they can be summarized in three main categories:
\begin{itemize}
\item "lost lepton": events with a lepton from a W boson decay, where the lepton is not identified. This background
comes from both W+jets and \ttbar+jets events.
\item  ÒirreducibleÓ: events that contain genuine \MpT~for example from Z+jets events, where the Z boson decays to neutrinos.
\item "fake  \MpT ": events that contain no real  \MpT, mostly QCD  multijet production  where the  \MpT~originates from instrumental mismeasurement.
\end{itemize}

\begin{figure}[htb]
\centering
\includegraphics[width=0.4\textwidth]{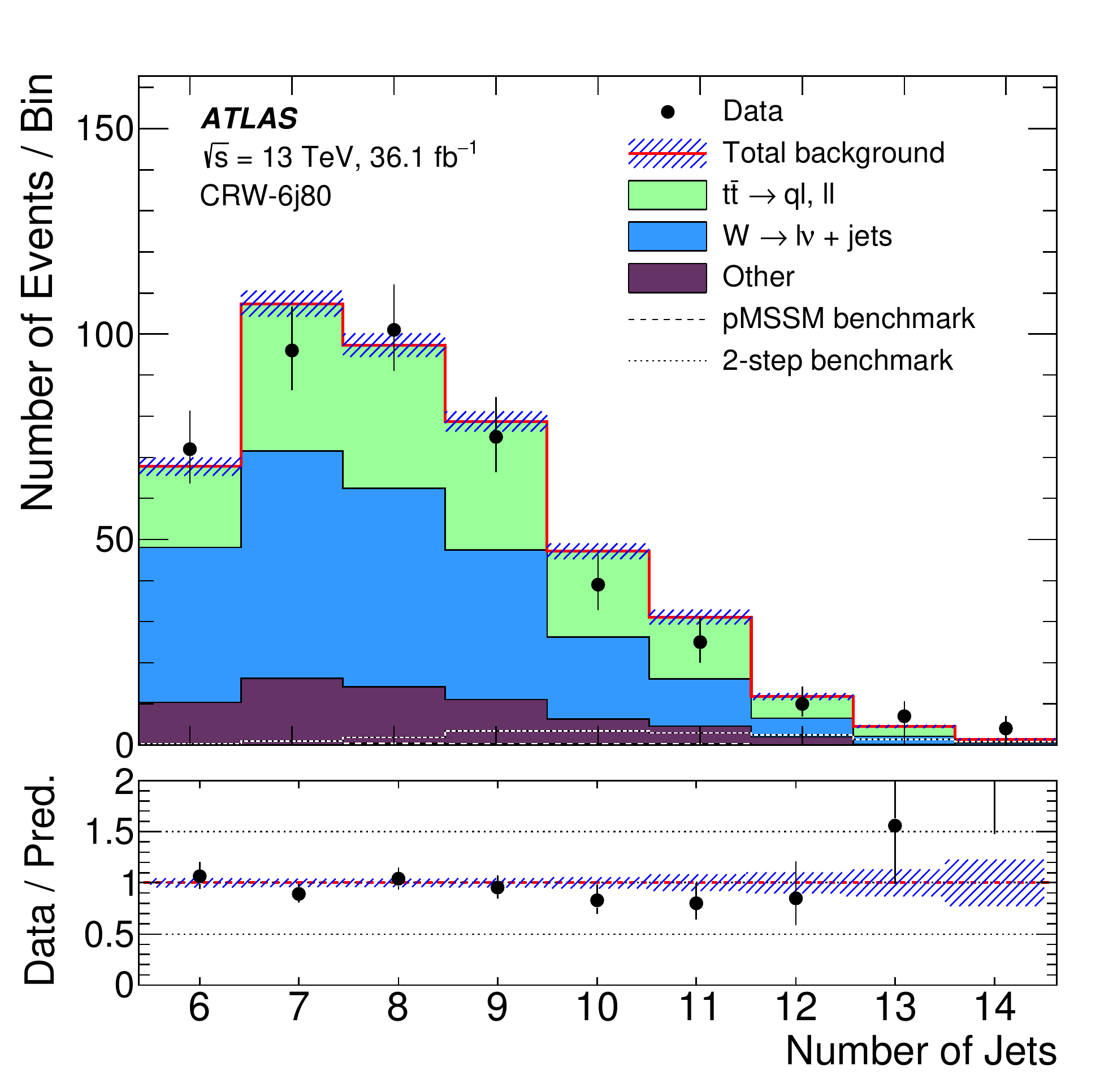}
\includegraphics[width=0.4\textwidth]{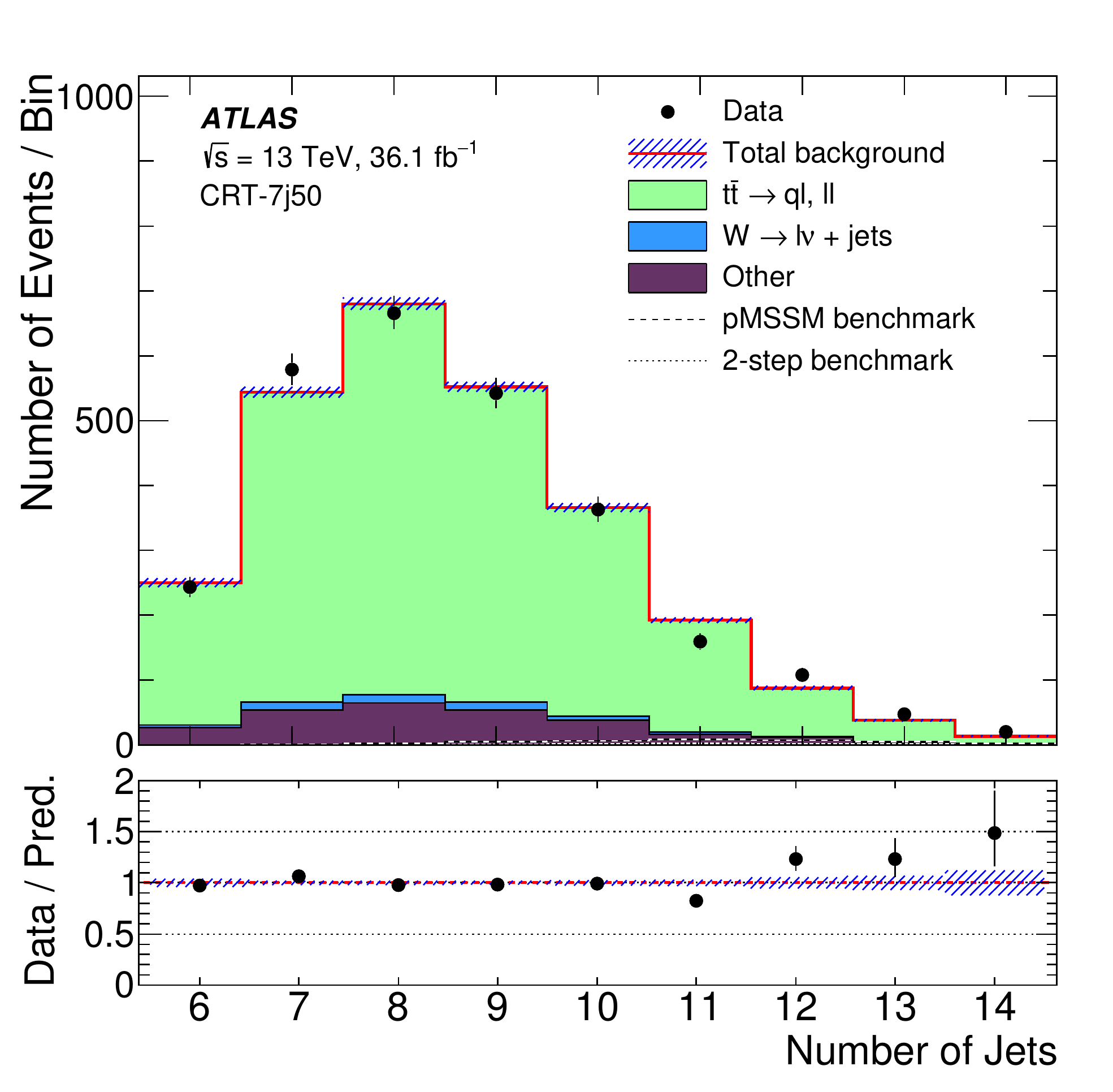}
\caption{Lost lepton W+jets and \ttbar~control region in Ref.~\cite{ATLAS-2017-33}. The \njet~distribution is shown for events containing exactly one lepton and zero b jets on the left (W+jets CR) and at least one b jet on the right (\ttbar~ CR).}
\label{fig:LLCR}
\end{figure}

These backgrounds can be determined with the help of simulation of known SM processes and definitions of control regions enhanced in the specific process of interest. For example, lost lepton backgrounds are estimated based on orthogonal regions where exactly one lepton is identified. In order to reduce the signal contamination in the control regions, events are selected where the transverse mass between the lepton and the \MpT~is below a threshold around the W boson mass. This control region is enhanced in W+jets and $t\bar{t}$+jets events, which can be further separated by requiring the absence of b jets or the presence of at least one b jet. Fig.~\ref{fig:LLCR} shows the \njet~distribution of these aforementioned control regions with zero b jets on the left and at least one b jet on the right for the ATLAS search in Ref.~\cite{ATLAS-2017-33}.
The normalization of the two background contributions is determined in these control regions.
Other control regions include the use of photon+jet events as proxy to estimate the background originating from Z boson decays into two neutrinos.
Additional validation regions, closer in terms of kinematic selection criteria to the signal regions, are often used to validate the different background estimation techniques.
Other data driven methods, such as rebalance-and-smear~\cite{CMS-SUS-16-033}, are employed to estimate QCD multijet production which include fake \MpT.

\subsection{Results and interpretation}
\begin{figure}[htb]
\centering
\includegraphics[width=0.7\textwidth]{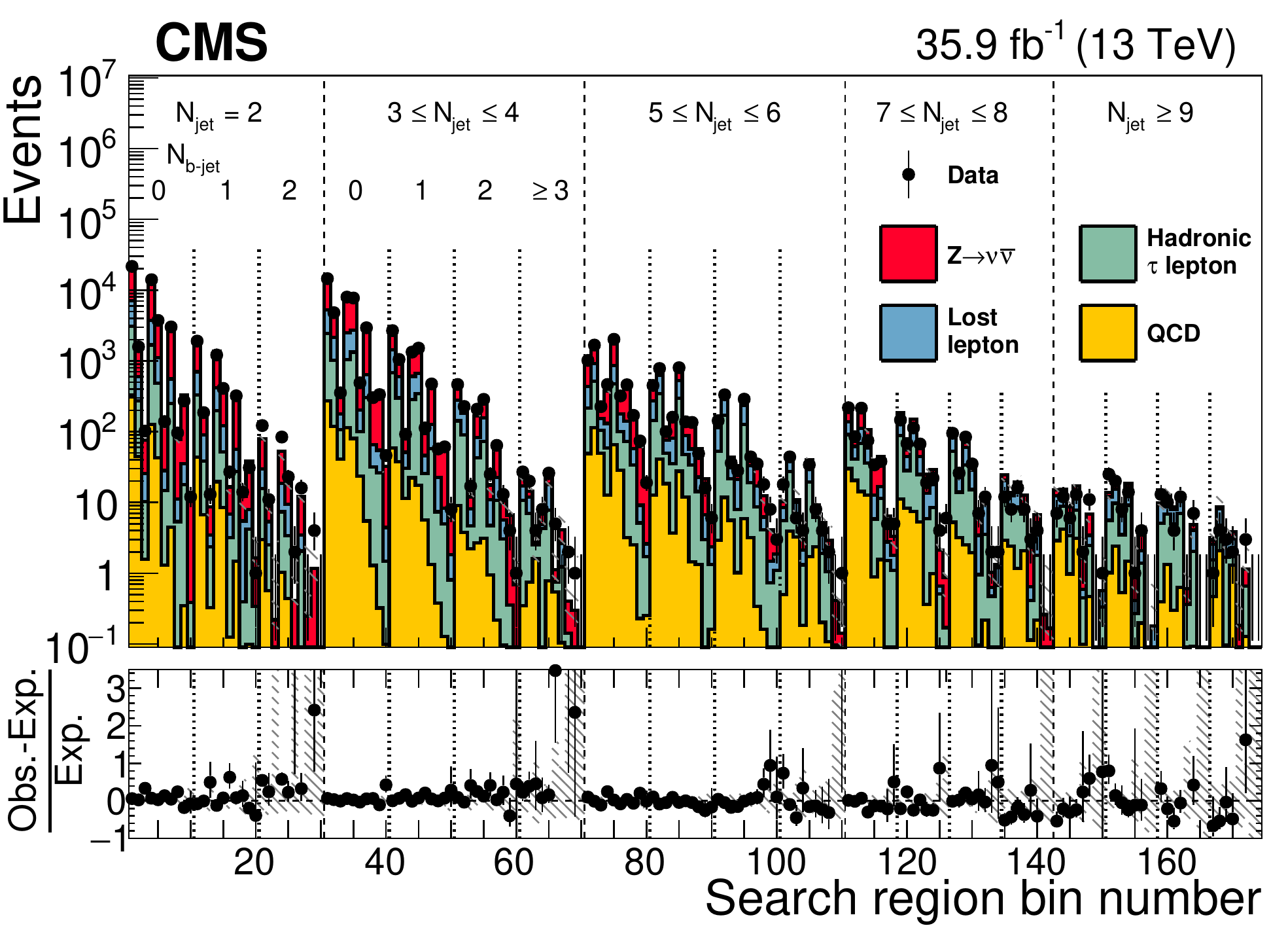}
\caption{The observed numbers of events and prefit SM background predictions in the 174 search regions of the analysis in Ref.~\cite{CMS-SUS-16-033}.}
\label{fig:result_16_003}
\end{figure}

The final result for each of the discussed searches compares the predicted number of background events and the observed number of events in each of the either statistically orthogonal search regions or the inclusively defined regions. 
Examples are shown in Fig.~\ref{fig:result_16_003} for one of the CMS searches~\cite{CMS-SUS-16-033}, where the observed number of events is shown in addition to the overall predicted number of background events separated by their origin. No significant deviations from SM backgrounds are observed for both CMS searches and limits are set on the production cross section of the previously discussed simplified models. The results are based on a statistical combination of all search regions. Limits at the 95 \% confidence level are set using a modified frequentist approach, employing the CL$_s$ criterion~\cite{Junk1999,ClsCite} and an asymptotic formulation~\cite{Cowan:2010js} of the profile likelihood ratio.

\begin{figure}[htb]
\centering
\includegraphics[width=0.4\textwidth]{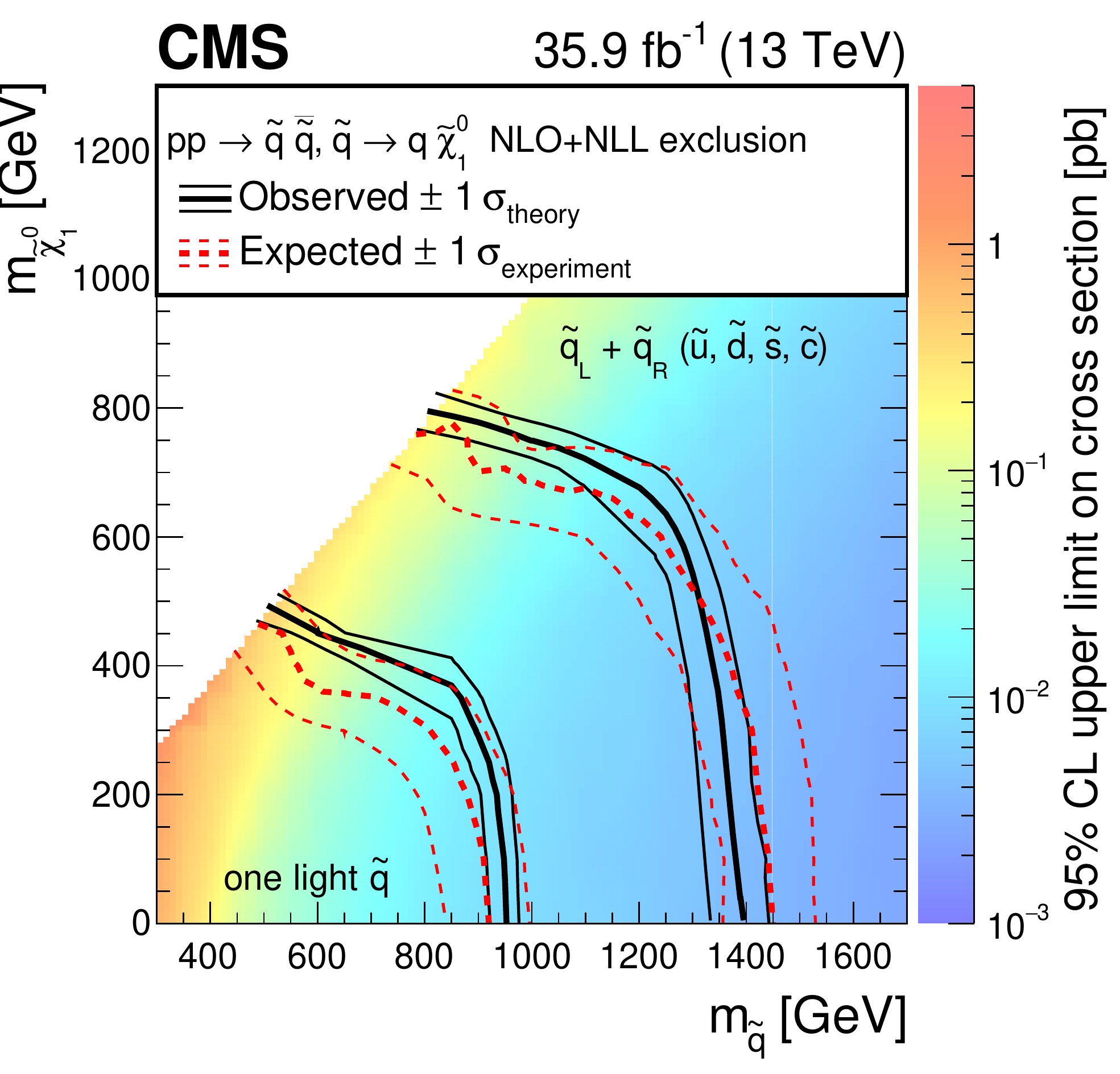}
\includegraphics[width=0.4\textwidth]{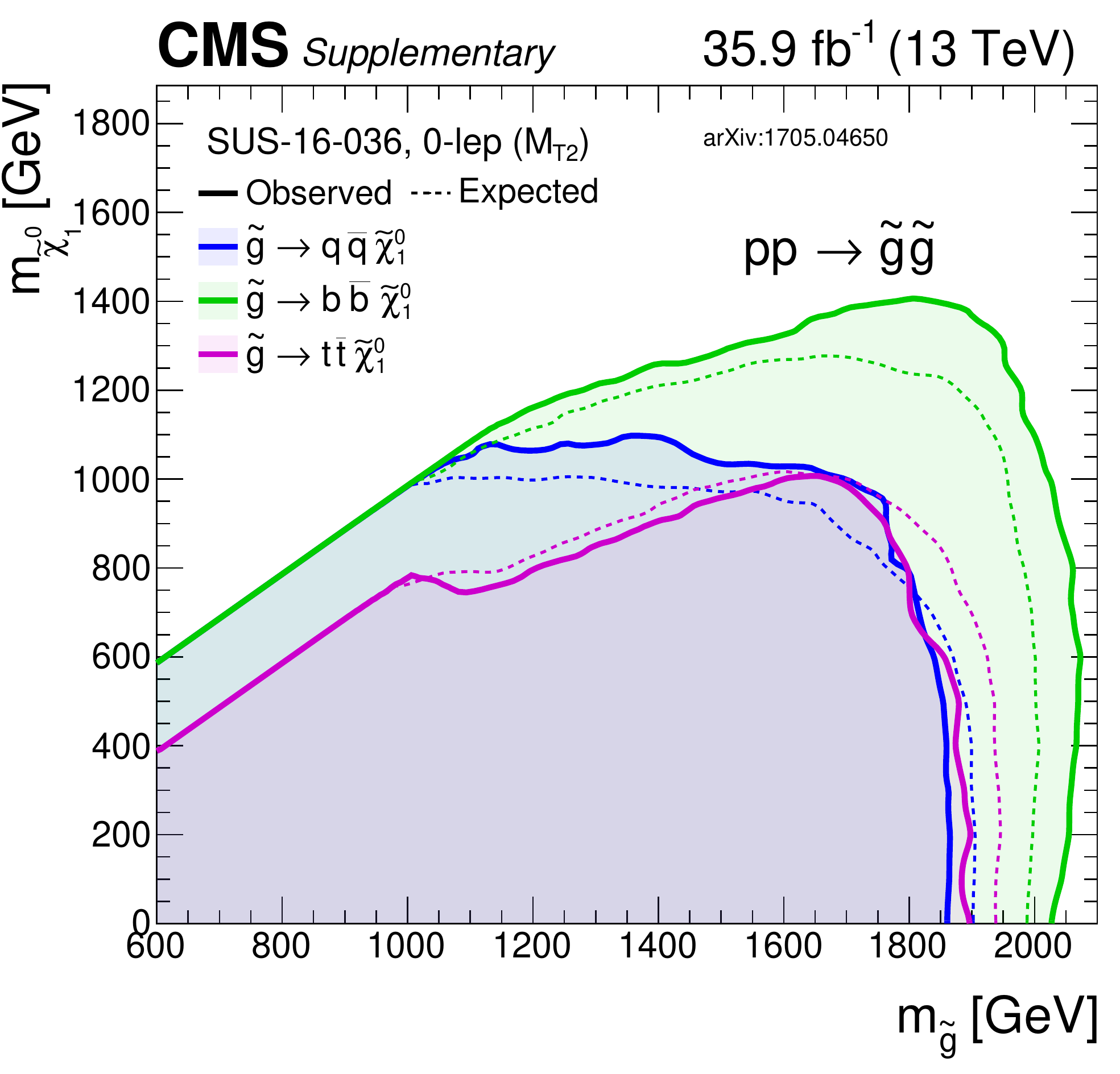}
\caption{The 95\% CL upper limits on the production cross sections for several simplified gluino models. The left plot shows the interpretation in Ref.~\cite{CMS-SUS-16-033} of a simplified model
where each gluino decays into two quarks and a neutralino. The right plot shows the results obtained in Ref.~\cite{CMS-SUS-16-036} for three different gluino decay scenarios.}
\label{fig:CMS_had_res}
\end{figure}

The left hand plot in Fig.~\ref{fig:CMS_had_res} shows the excluded phase space of direct light flavor squark production in the squark - LSP plane. The excluded squark mass reaches up to about 1400 GeV for low neutralino masses. The right plot in the same figure shows a summary of the results from Ref.~\cite{CMS-SUS-16-036} for gluino pair production in the different decay modes to either a light flavor quark pair, a bottom quark pair, or a top quark pair and the LSP. The high mass limits extend up to 1850 GeV, 1875 GeV, and 2000 GeV for the respective models and low LSP masses.

\begin{figure}[htb]
\centering
\includegraphics[width=0.5\textwidth]{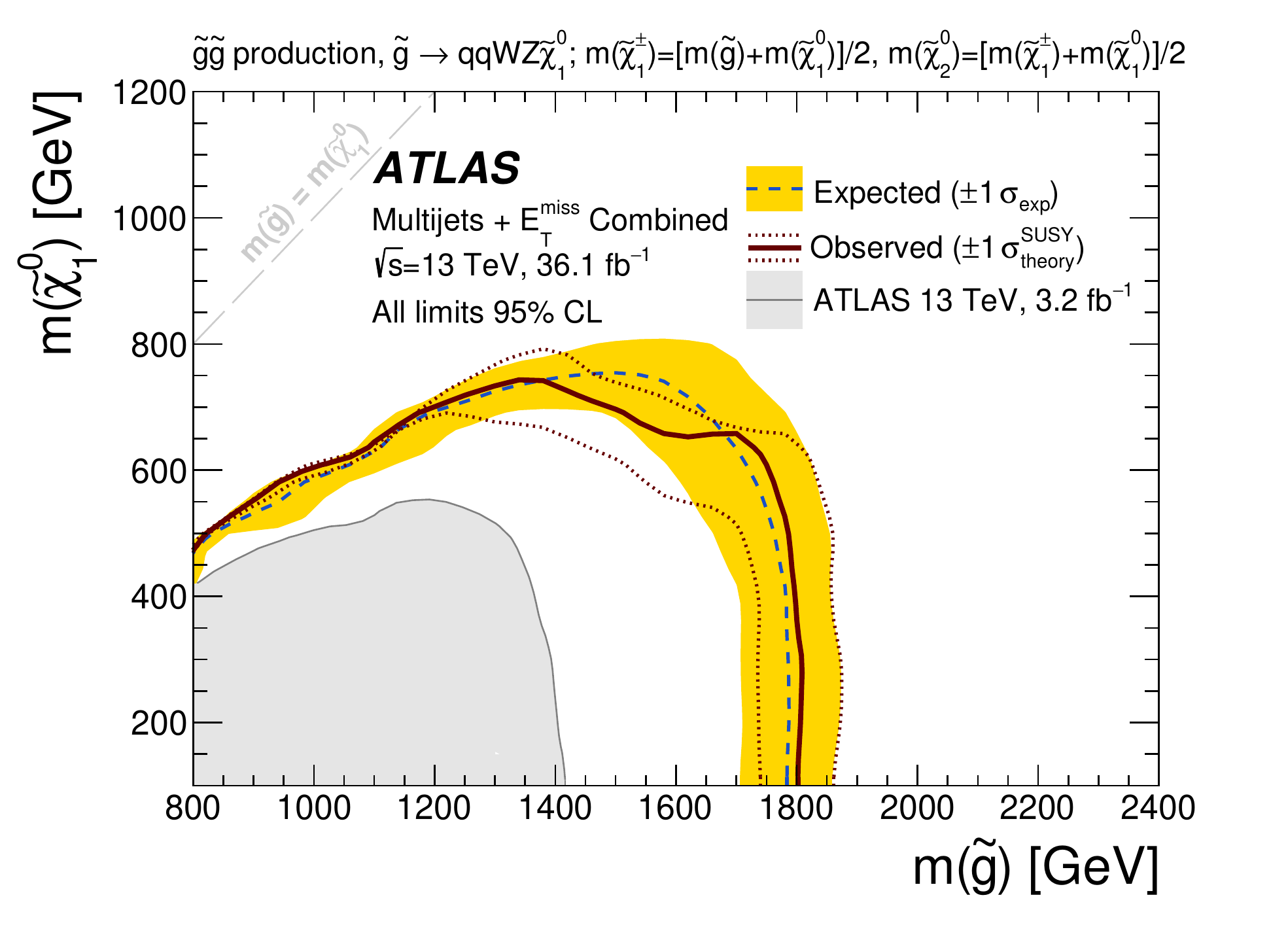}
\caption{Interpretation of the result from Ref.~\cite{ATLAS-2017-33}. The plot shows the exclusion limit for a gluinos with a two step cascade decay wich contains W and Z boson.}
\label{fig:result17_033}
\end{figure}

The plot in Fig.~\ref{fig:regionresult17_033} shows the \MpTHT~distribution for one of the search regions with at least 10 jets and $MJ > $500 GeV of the high jet multiplicity ATLAS search~\cite{ATLAS-2017-33}. Good agreement is observed between the predicted and observed event yields. For each point in the two dimensional phase space, depicted in Fig.~\ref{fig:result17_033}, the search region with the best expected limit is used. The model refers to pair produced gluinos with a two step cascade decay wich contains
W and Z bosons in the final state as shown in the diagram in Fig.~\ref{fig:feynman}. Gluino masses up to 1800 GeV are excluded for low LSP mass in this scenario.

\section{Searches with up to one lepton}
\subsection{Analyses strategy}
In addition to purely hadronic final states many of the already discussed models result in  leptons in the final state.
The simplified model shown on the top right of Fig.~\ref{fig:feynman} leads to four top quarks in the final state which leads to about a 40 \% probability of having
exactly one lepton ($e$, $\mu$, $\tau$) in the end through the decays of the W bosons. Additionally, gluino pair production that contains top and bottoms quarks in the decay can be probed
as well through the identification of the these bottom quarks.
Two separate CMS analyses~\cite{CMS-SUS-16-037, CMS-SUS-16-042} target the four top final state, while the latter additional contains a separate search for final states with four light flavor quarks and two W bosons.
The ATLAS search described in~\cite{ATLAS-17-21}  uses a combination all hadronic and single lepton final states to target simplified models with four top or bottom quarks.

For the CMS search described in Ref.~\cite{CMS-SUS-16-037} events are selected that contain exactly one lepton and the previously discussed $MJ$ variable is used, where small radius $R=0.4$ jets and the lepton are reclustered into larger radius $R=1.4$ jets, whose masses are added up yielding the final value of $MJ$ for each event. Additional requirements on the transverse mass $M_T(l,\MpT)$ to be above 140 GeV define the phase space of the search, which is then further subdivided into different bins of \MpT, the number of jets, and b jets (with the requirement of at least one b jet in the final state).
Ref.~\cite{CMS-SUS-16-042} describes a search with a similar baseline selection of one lepton, however here requirements are placed on \LT~the sum of the lepton \pT~and \MpT~which
represents the \pT~of the W boson in SM events. The main discriminating variable in the search is defined as the azimuthal angle between the reconstructed W boson (based on \MpT~and the lepton) and the lepton itself $\Delta \phi(W,l)$. This variable shows a steeply falling distribution for SM events with one W boson decaying into a lepton and a neutrino, however it is essentially flat for SUSY processes.
Search regions are defined based on the number of jets, b jets, \LT~and \HT, with a varying selection on $\Delta \phi(W,l)$ depending on the \LT~category.

The ATLAS search described in~\cite{ATLAS-17-21} utilizes a baseline selection with at least three b jets, and either zero or one lepton in the final state.
Two different approaches are chosen, where either inclusive search regions are defined based on $MJ$, \meff,  \MpT, and $M_T$, or where exclusive
search bins are used and categories are defined based on the number of jets and \meff. 

\subsection{Background modeling and interpretation}
\begin{figure}[htb]
\centering
\includegraphics[width=0.3\textwidth]{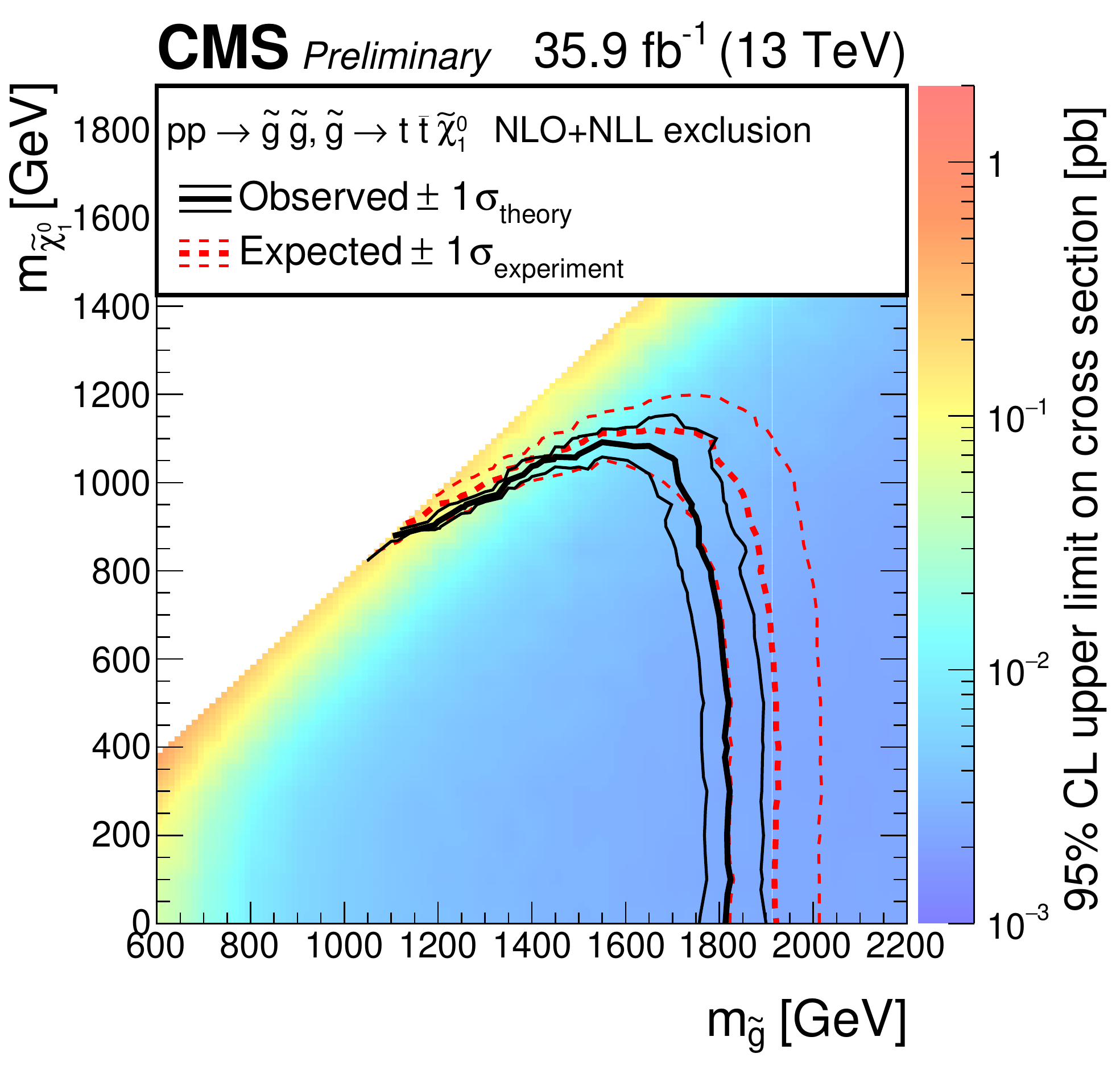}
\includegraphics[width=0.3\textwidth]{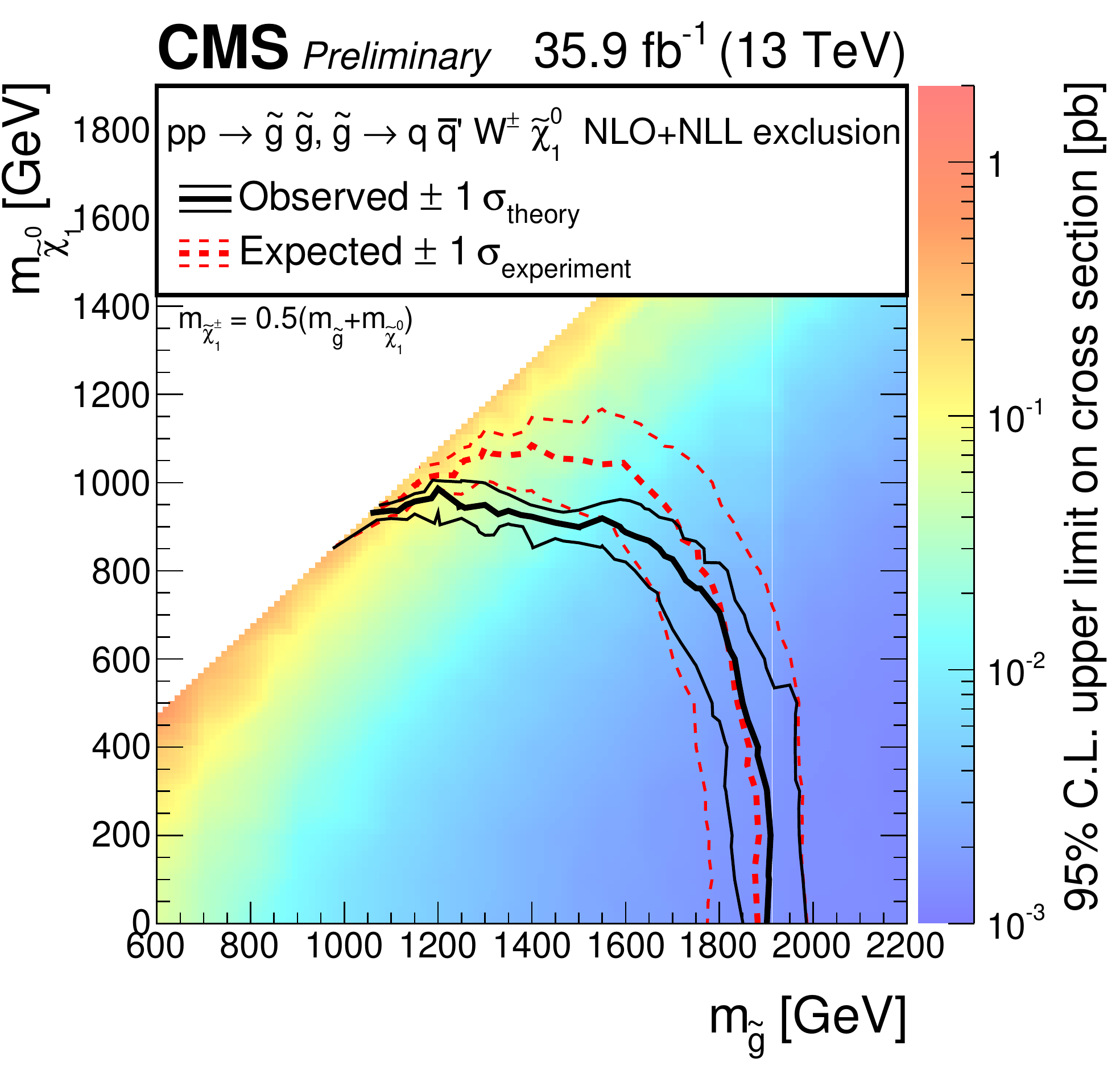}
\includegraphics[width=0.3\textwidth]{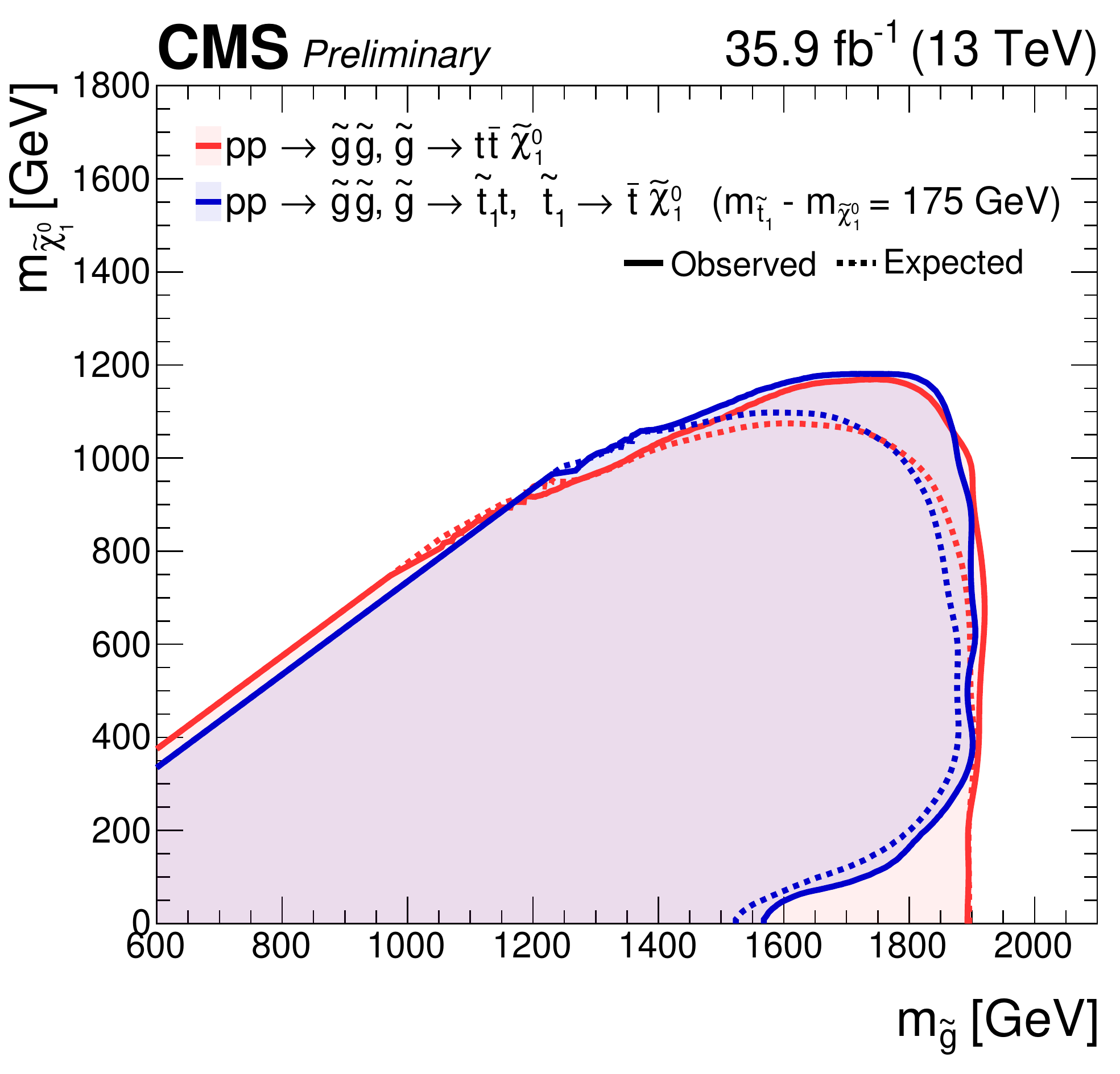}
\caption{The 95\% CL upper limits on the production cross sections for several simplified gluino models from Ref.~\cite{CMS-SUS-16-037} (left) and Ref.~\cite{CMS-SUS-16-042} (middle, right).}
\label{fig:CMS_1l_result}
\end{figure}

The remaining SM background for the two CMS searches originate mainly from dileptonic \ttbar~ events, where one lepton was lost, which is estimated in data with the help of side band and control regions, similar to an ABCD method.
Correction factors extracted from simulation are then used to predict the background in the signal regions. Both searches exhibit a similar reach in sensitivity for the gluino pair production model with four top quarks
in the final state, as shown in the left and middle plot in Fig.~\ref{fig:CMS_1l_result}, reaching up to about 1900~GeV. The right plot in the same figure shows the mass exclusion limit for the model containing four light
flavor quarks and two W bosons, reaching a mass exclusion of up to 1900~GeV as well.

\begin{figure}[htb]
\centering
\includegraphics[width=0.7\textwidth]{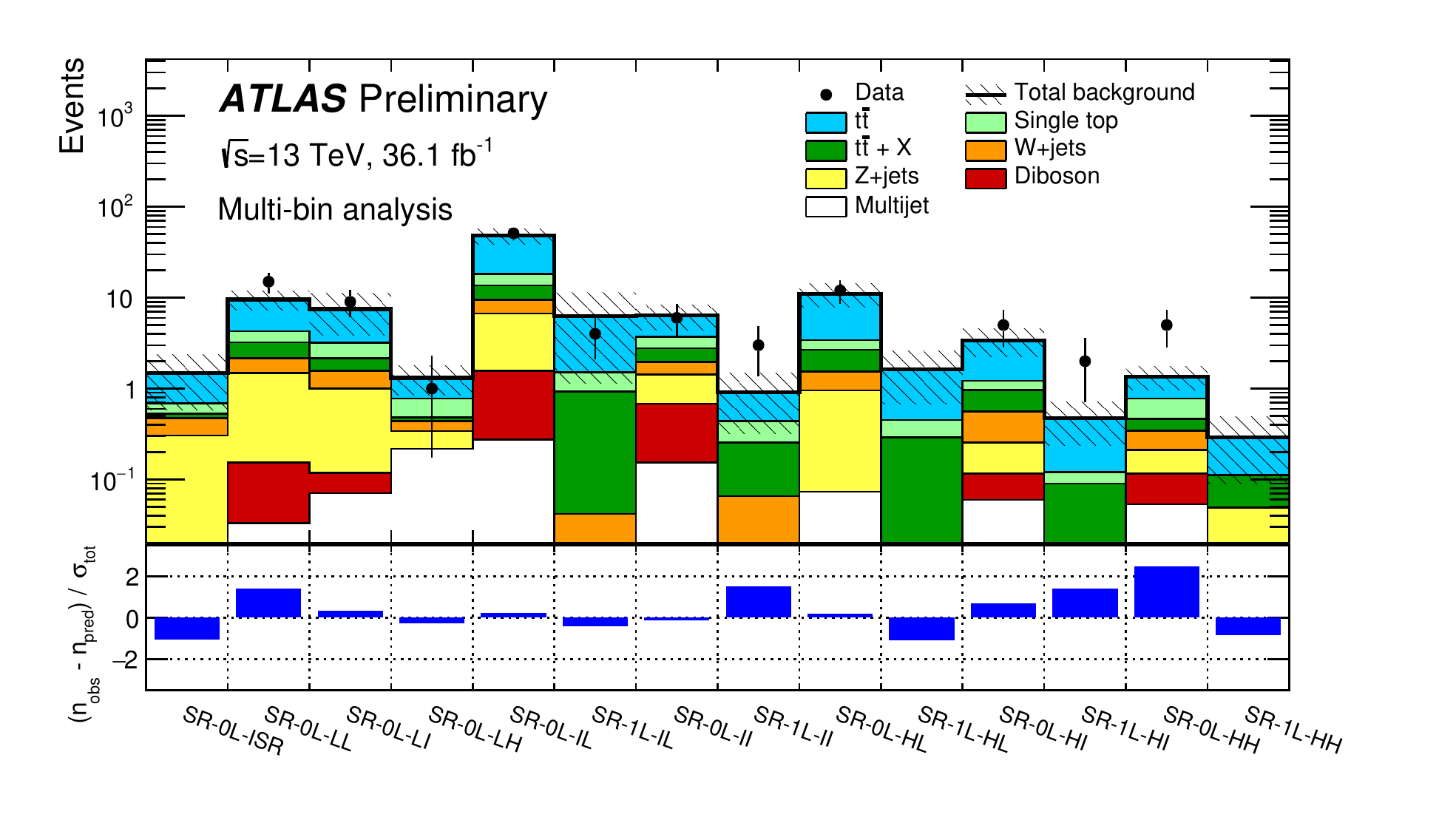}
\caption{The observed numbers of events and SM background predictions in the exclusive search regions of the analysis in Ref.~\cite{ATLAS-17-21}.}
\label{fig:ATLAS_1l_regions}
\end{figure}

\begin{figure}[htb]
\centering
\includegraphics[width=0.4\textwidth]{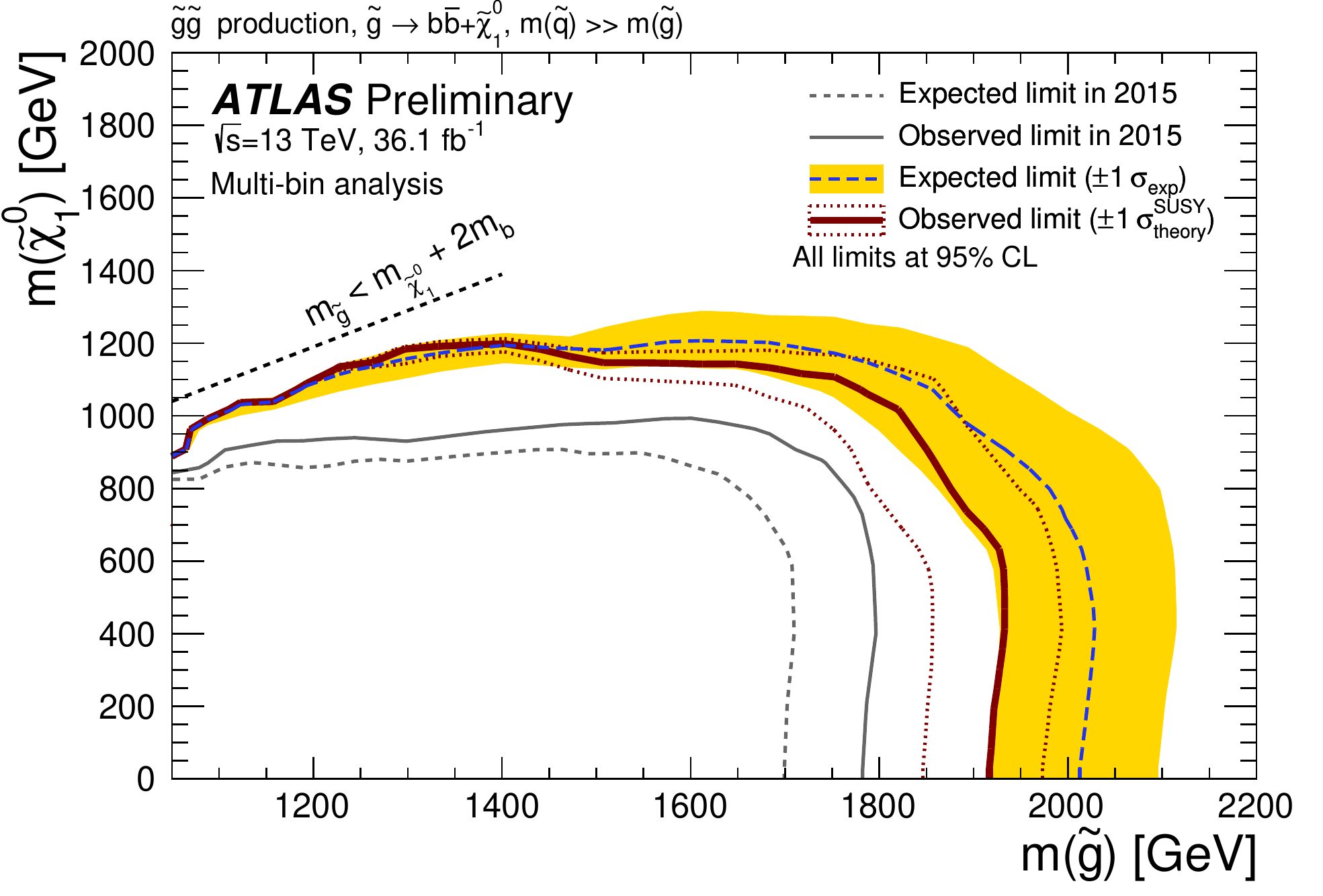}
\includegraphics[width=0.4\textwidth]{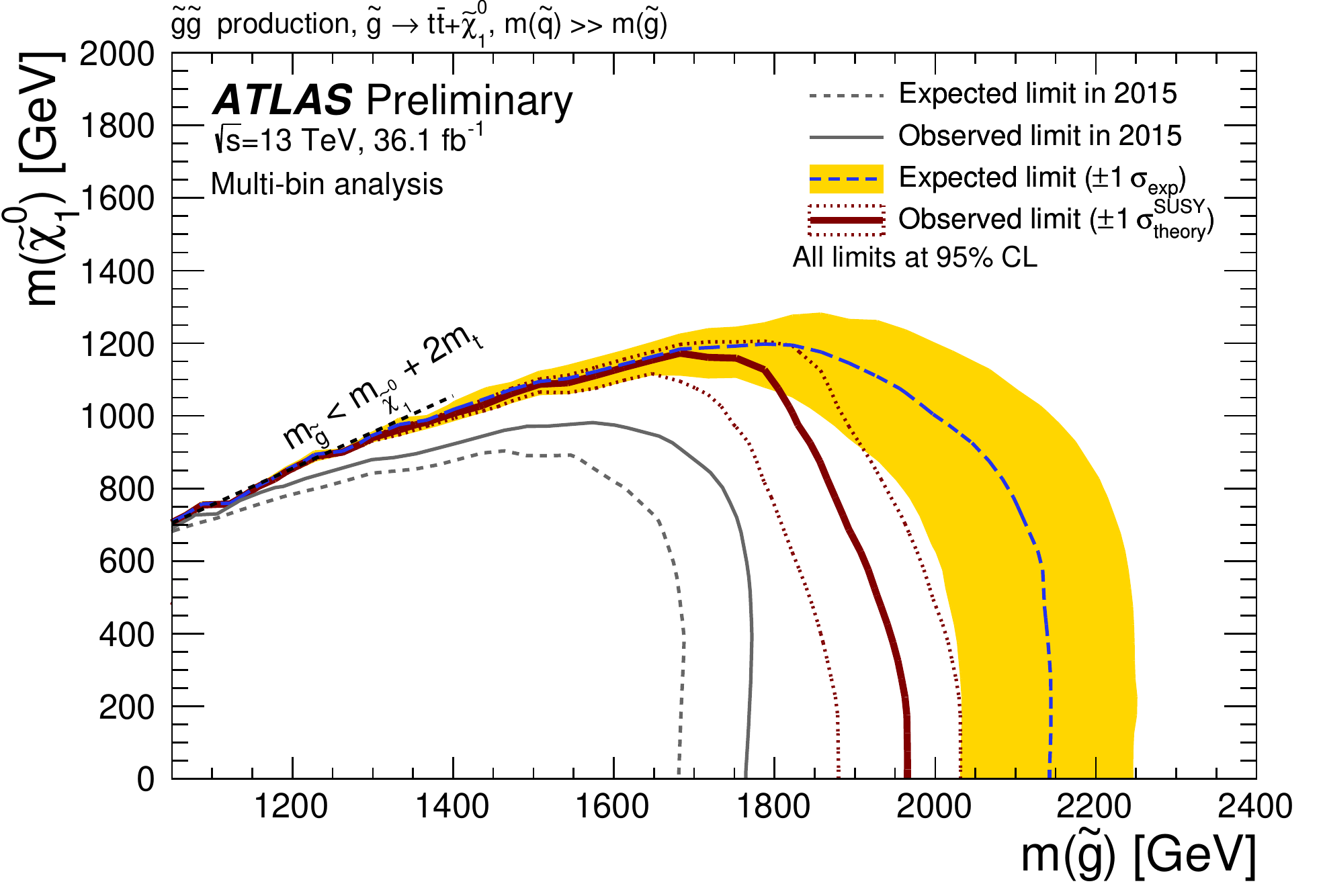}
\caption{The 95\% CL upper limits on the production cross sections for several simplified gluino models from Ref.~\cite{ATLAS-17-21}. Gluino decays to two b quarks and a neutralino
are shown on the left, while gluino decays to two top quarks and a neutralino are shown on the right.}
\label{fig:ATLAS_1l_results}
\end{figure}

For the ATLAS search the normalization of the dominant \ttbar~background is determined in orthogonal control regions, and 
validated in dedicated validation regions. The final result of the expected background and the observed number of events in the 10 exclusive signal regions is shown in Fig.~\ref{fig:ATLAS_1l_regions}.
Good agreement is found and Fig.~\ref{fig:ATLAS_1l_results} shows the excluded mass scenarios as a function of gluino versus LSP mass for two different decay models of pair produced gluinos.
The left hand side contains limits for the scenario where each gluino decays into two b quarks and the neutralino, where a mass exclusion of up to 1900 GeV is reached for low LSP masses. The right hand side shows exclusion limits for the scenario where each gluinos decays into two top quarks and the LSP, reaching up to 1950 GeV for low LSP masses.
In both scenarios no intermediate particles are considered and squark masses are assumed to be much higher than the gluino masses.

\section{Multilepton searches}
 \subsection{Analysis strategy}
 
 While multilepton final states are often associated with relatively small branching ratios in the strongly produced SUSY models considered here, they also offer a very clean environment with lower background yields
 than the previously discussed all hadronic and single lepton searches. The ATLAS search described in Ref.~\cite{ATLAS-2017-030} makes use of a same-sign dilepton or three lepton selection, while the focus of the CMS search described in Ref.~\cite{CMS-SUS-16-035} is on the same-sign dilepton final state.
 Processes that produce two leptons with the same charge are especially rare in the SM, compared to the production of oppositely charged leptons.
 In the ATLAS search, inclusive signal regions are defined based on the number of leptons, with a minimum requirement of at least two leptons with a \pT~ $>$ 20 GeV, or with the trailing lepton \pT~ $>$ 10 GeV.
 Additional selection requirements are placed on the number of jets and b jets with varying \pT~thresholds, \MpT, \meff, and \MpT/\meff. Several of the search
 regions target $R$-parity violating processes and are not discussed in detail here.
 The CMS search categorizes events based on the \pT~of the two leptons in the event with the same charge. 
 The two \pT~categories for leptons are either labeled as Low (L) for 10 GeV $<$ \pT~$<$ 25 GeV or High (H) for \pT~$>$ 25 GeV, and events are placed into three categories HH, HL, or LL. Additional exclusive search regions are defined based on the number of jets, the number of b jets, \HT~and \MpT.

 \subsection{Background estimation}
 
\begin{figure}[htb]
\centering
\includegraphics[width=0.7\textwidth]{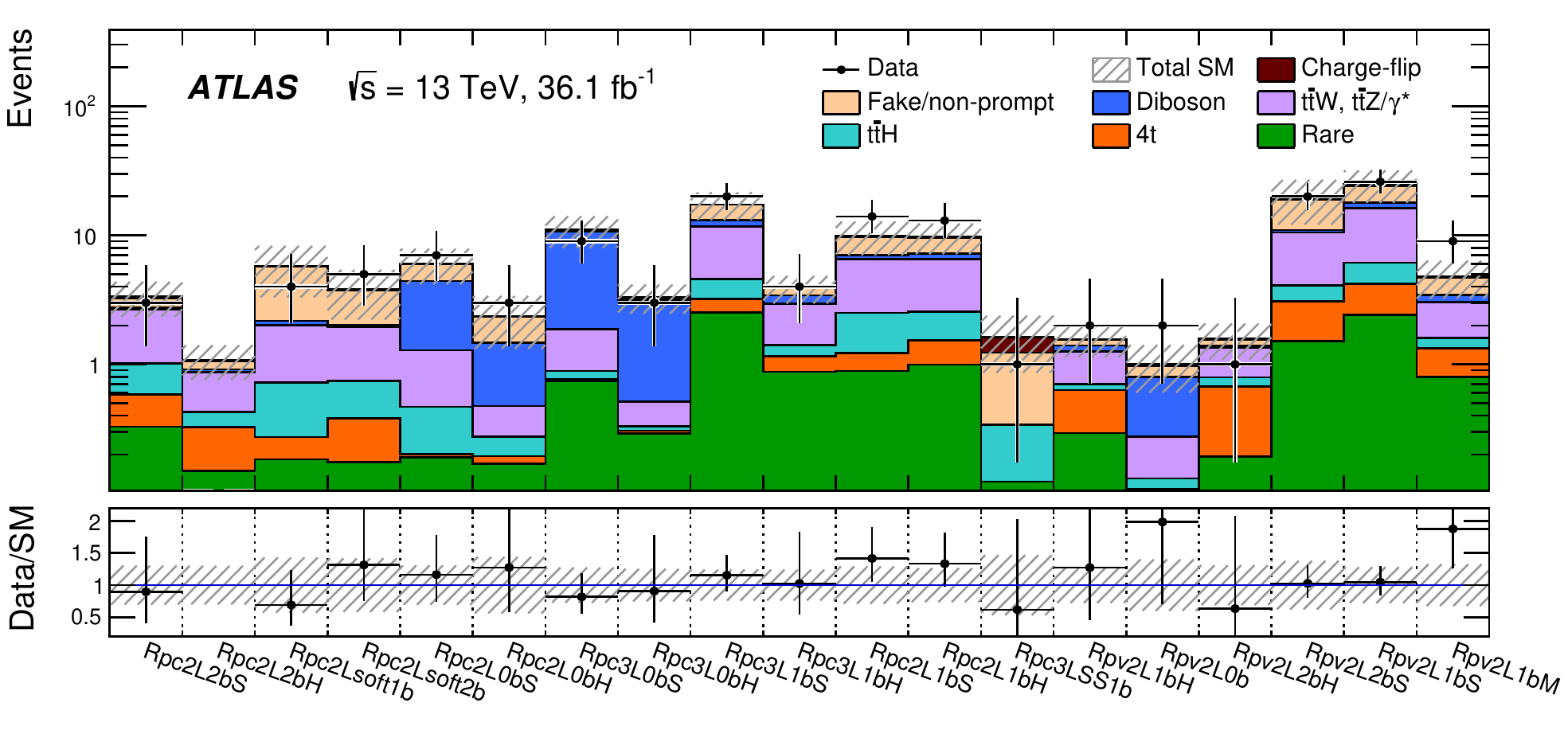}
\caption{The observed numbers of events and SM background predictions in the search regions of the analysis in Ref.~\cite{ATLAS-2017-030}.}
\label{fig:ATLAS_multilept_bins}
\end{figure}

 The main irreducible backgrounds for this type of event selection arise from \ttbar$V$ and diboson production, which are estimated from simulation, corrected in control regions, and the prediction tested in dedicated validation regions.
 Reducible backgrounds contribute similarly to the expected background composition in the signal region. The dominant ones arise from the misidentification of the lepton charge due to the interaction of the lepton with the tracker material and the emission of hard bremstrahlung electrons. These electrons can lead to an asymmetric production of an 
 electron-positron pair, where only one of them has a large enough \pT~to be identified.
 Additionally, non-prompt or fake leptons are another source of background. These contributions are estimated via a matrix method, making use of tight and loose lepton selection criteria and estimating the probability of a loosely identified lepton to also pass the tight selection.
  Good agreement is found between the expected SM backgrounds and the observed yields in the different signal regions of the CMS and ATLAS searches.
 Fig.~\ref{fig:ATLAS_multilept_bins} shows the result of the inclusive search categories of the ATLAS analysis.

\begin{figure}[htb]
\centering
\includegraphics[width=0.3\textwidth]{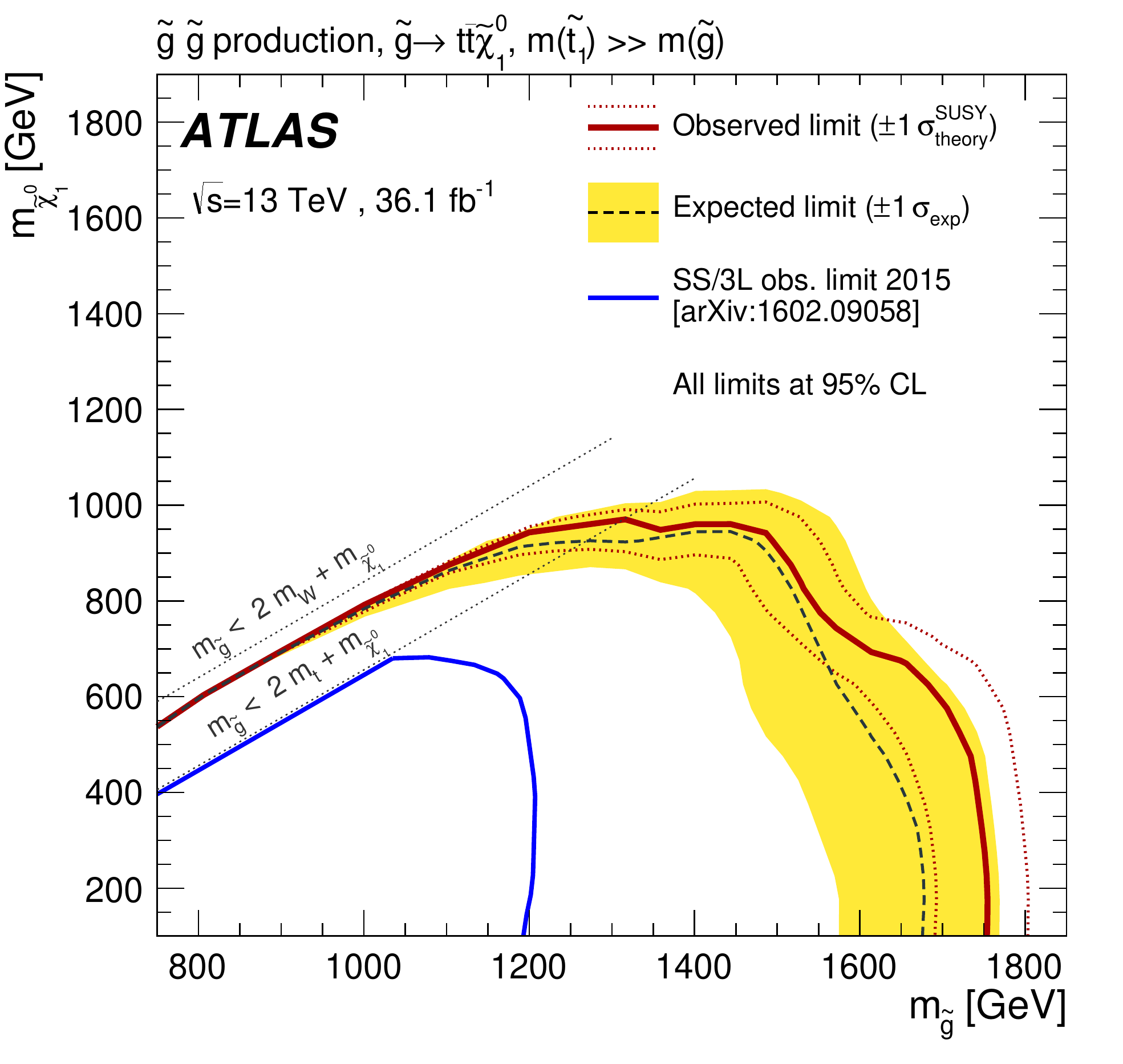}
\includegraphics[width=0.3\textwidth]{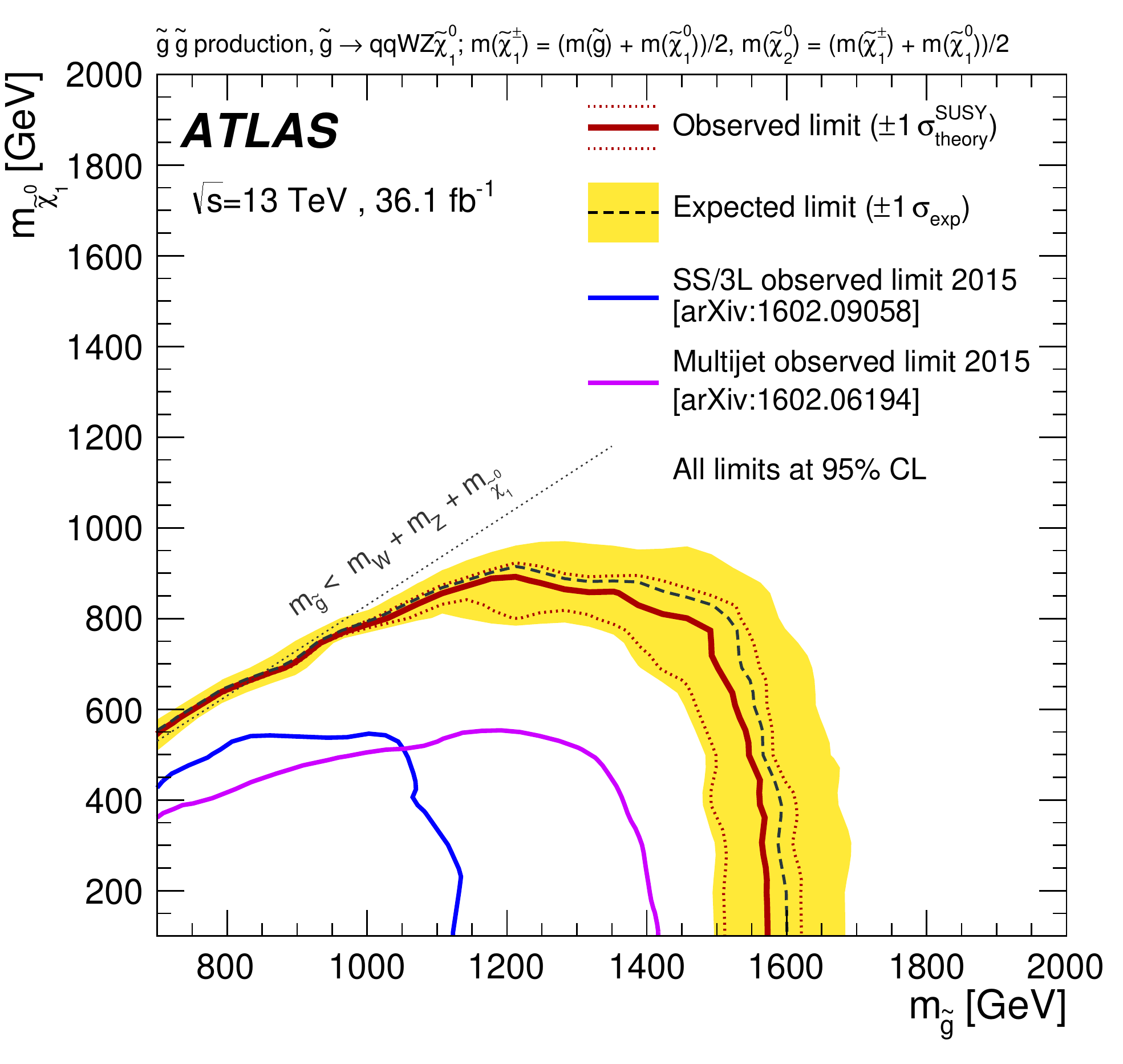}
\includegraphics[width=0.3\textwidth]{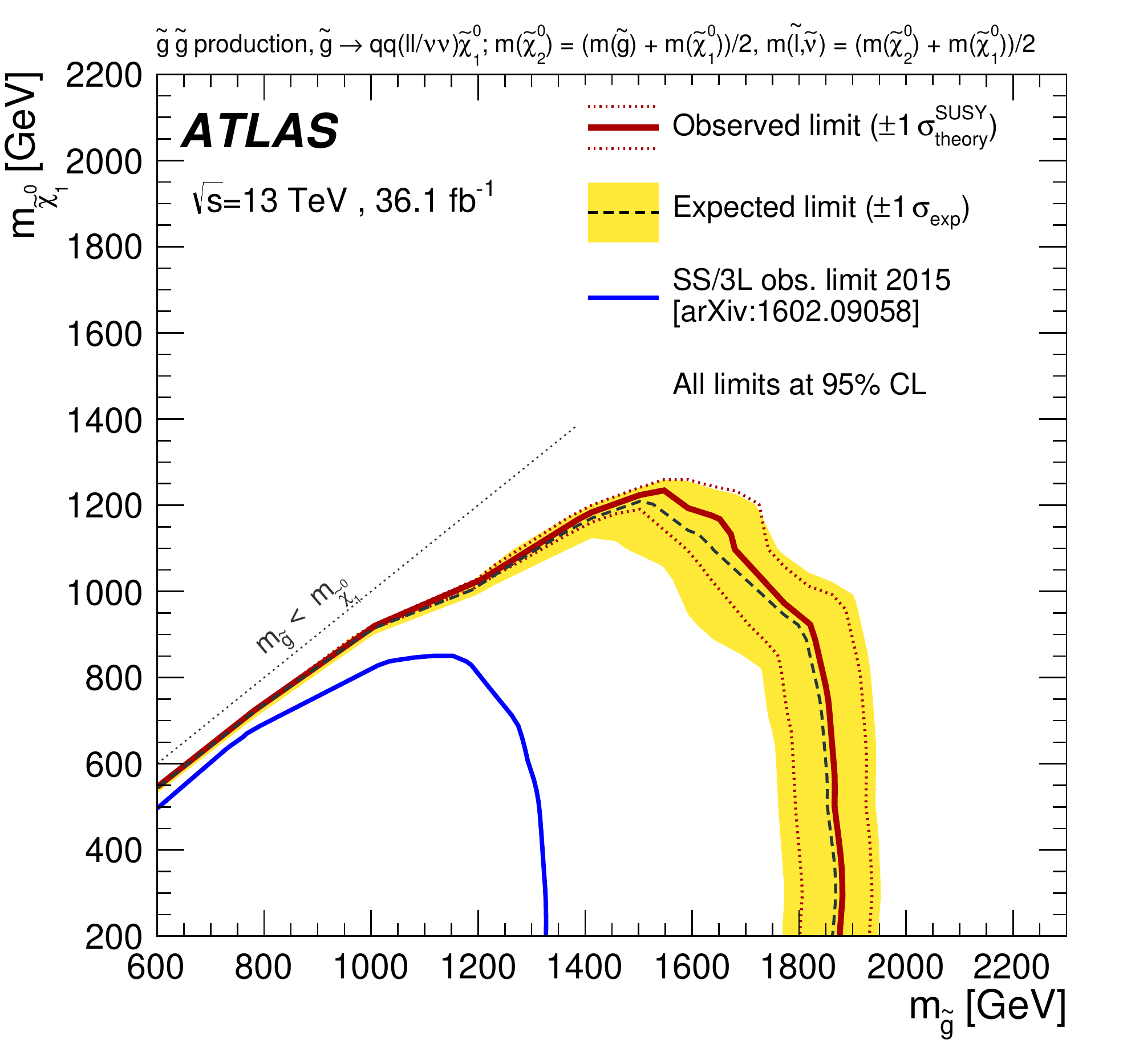}

\caption{The 95\% CL upper limits on the production cross sections for several simplified gluino models from Ref.~\cite{ATLAS-2017-030}.
The left most plot shows the limit on gluino pair production in a four top final state, while the middle and right plot show limits for two step cascade processes.}
\label{fig:ATLAS_multilept_limit}
\end{figure}

 The results are interpreted in a variety of different final states including several cascade decays with intermediate W and Z bosons as well as other models with sleptons, leptons, and neutrinos.
 Examples from ATLAS are shown in Fig.~\ref{fig:ATLAS_multilept_limit}. The left most plot shows the limit on gluino pair production in a four top final state, reaching
 up to about 1700 GeV. The middle and right plot show limits for two step cascade processes, where the observed excluded mass reaches to about 1600 GeV for decays involving
 charginos and W or Z bosons and 1800 GeV for decays involving sleptons. The blue and purple lines indicate the observed exclusion limits with the 2015 dataset at 13 TeV, showing improvements of 
 several hundred GeV for high gluino and low LSP masses.
 
\begin{figure}[htb]
\centering
\includegraphics[width=0.4\textwidth]{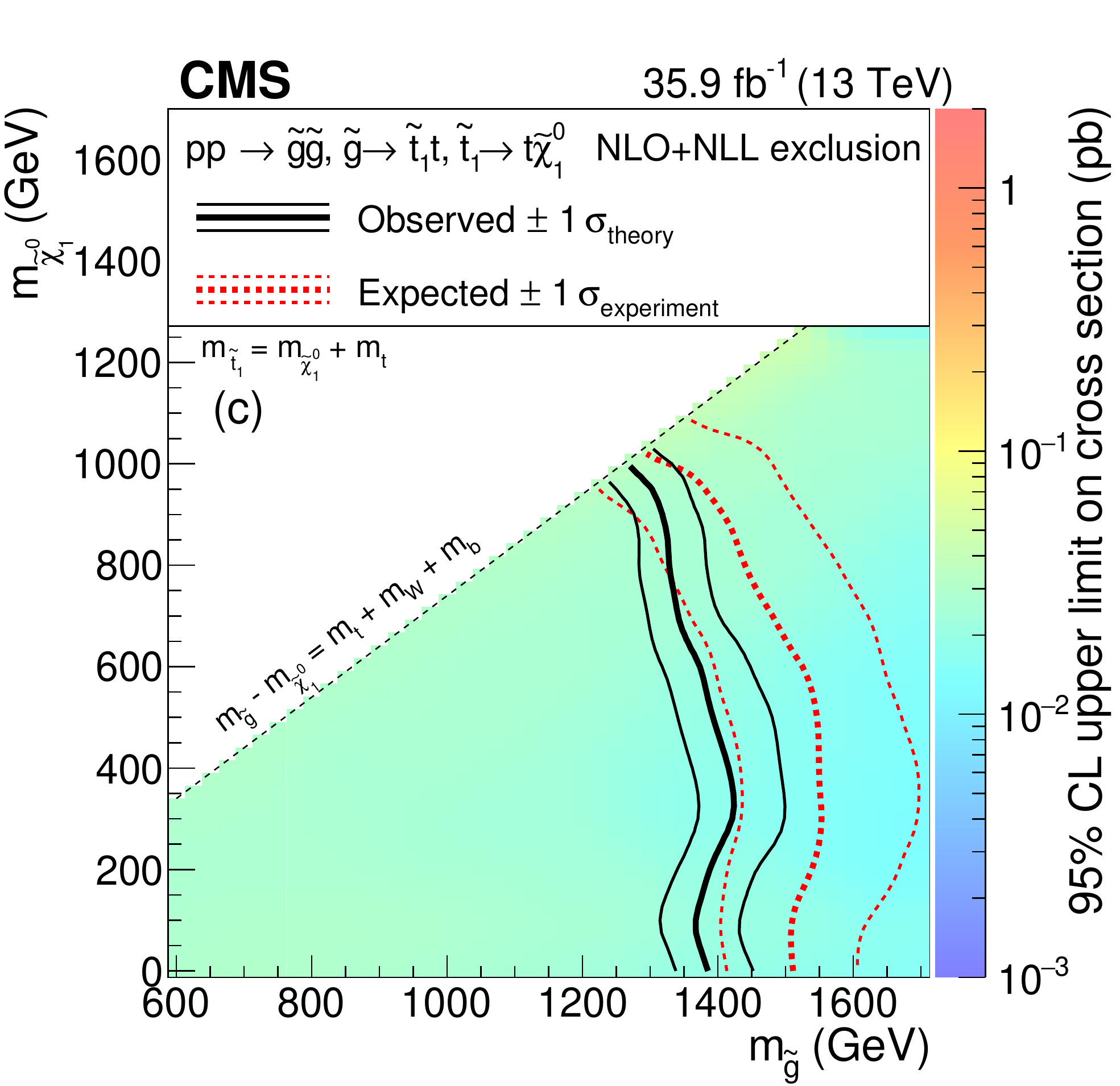}
\includegraphics[width=0.4\textwidth]{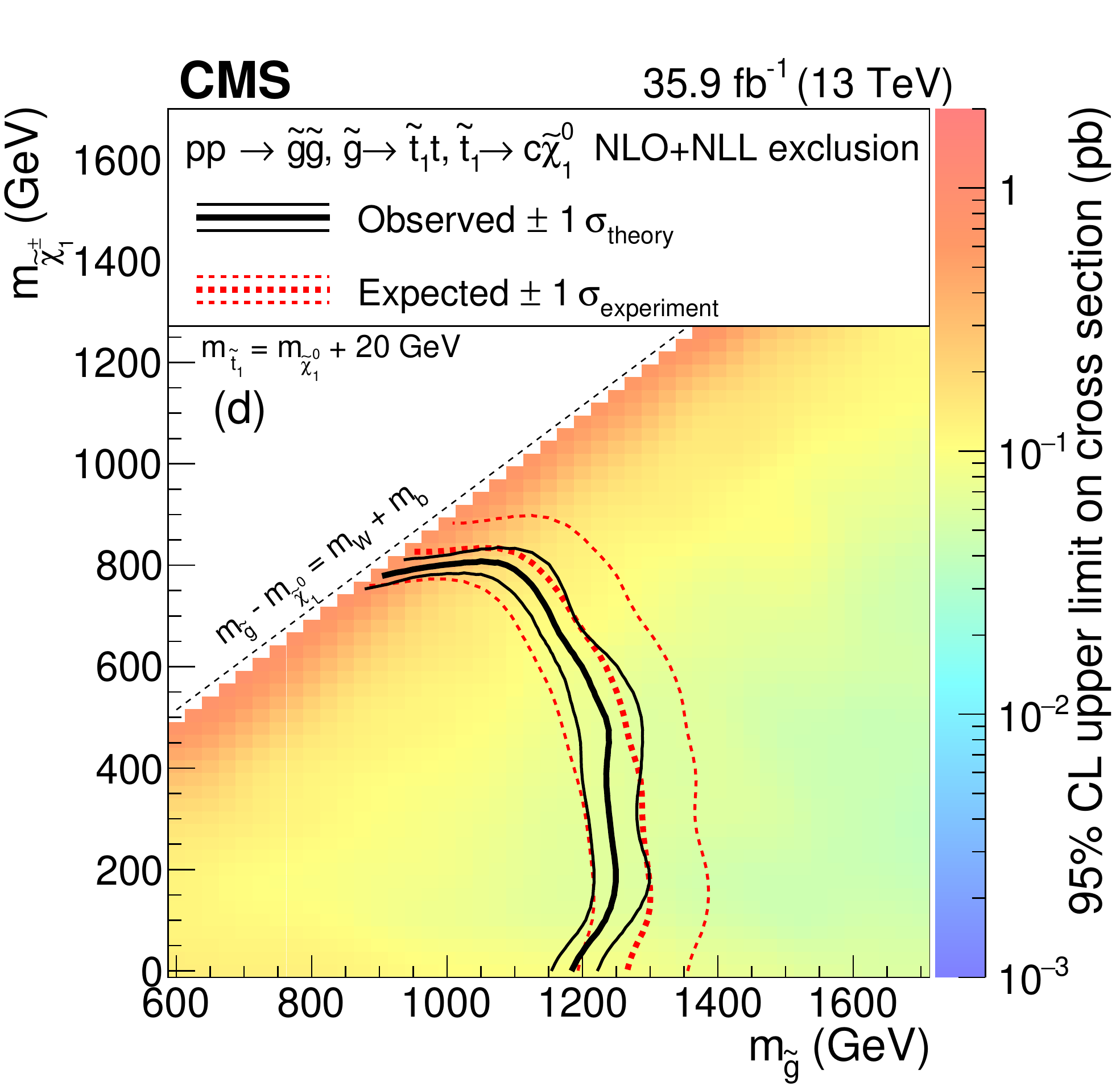}
\caption{ The 95\% CL upper limits on the production cross sections for several simplified gluino models from Ref.~\cite{CMS-SUS-16-035}. Both interpretations
consider one step cascade decays with intermediate top squarks.}
\label{fig:CMS_multilept_limit}
\end{figure}

 Two CMS interpretations are shown in Fig.~\ref{fig:CMS_multilept_limit} describing cascade decays of gluinos into either a four top final states, where the intermediate top squark is 
 set to the neutralino mass plus the top quark mass, or through a flavor changing neutral current (fcnc) process into a two top quark, two charm quark and LSP final state.

\section{Conclusion}
Both experiments, ATLAS and CMS, have an extensive search program covering strongly produced squark and gluino scenarios.
A variety of different final states has been analyzed making use of the full dataset of 36 fb$^{-1}$ collected during 2016. Overall
good agreement is found between the expected standard model backgrounds an the observed number of events.
The results shown in these proceedings cover interpretations of the experimental results in the form of limits on the production cross section of a variety of simplified SUSY models.
Limits on gluino pair production processes reach up to 2000 GeV for low LSP masses, depending on the decay channel of the gluino.
 \clearpage



\bibliographystyle{myieeetr}

\bibliography{bibfileFull}{}

\end{document}

%% file: econfmacros.tex
\def\beq{\begin{equation}}
\def\eeq#1{\label{#1}\end{equation}}
\def\eeqn{\end{equation}}

\def\beqa{\begin{eqnarray}}
\def\eeqa#1{\label{#1}\end{eqnarray}}
\def\eeqan{\end{eqnarray}}

\let\bar=\overbar

\def\Dslash{\not{\hbox{\kern-4pt $D$}}}
\def\dslash{\not{\hbox{\kern-2pt $\del$}}}

\def\msb{{\bar{\ssstyle M \kern -1pt S}}}

